\documentclass[10pt]{article}

\usepackage{epsfig}

\newlength{\dinwidth}
\newlength{\dinmargin}
\def\g{^{\circ}}
\newcommand{\lsim}{\lower .7ex\hbox{$\;\stackrel{\textstyle <}{\sim}\;$}}
\newcommand{\gsim}{\lower .7ex\hbox{$\;\stackrel{\textstyle >}{\sim}\;$}}

\begin{document}
\titlepage

\vspace*{2cm}

\begin{center}

{\Large \bf Correlation between regions of star formation and gamma-ray sources}

\vspace*{1cm}
\textsc{K. Belotsky, A. Galper, B. Luchkov} \\

\vspace*{0.2cm}
Moscow Engineering Physics Institute,
Moscow, Russia \\[0.5ex]
%$^c$ Center for Cosmoparticle Physics "Cosmion" of
%Keldysh Institute of Applied Mathematics, \\
%Moscow, 125047, Russia \\[0.5ex]

\end{center}

\vspace*{1cm}

\begin{abstract}
Statistically significant correlation between positions of unidentified gamma-ray sources (UGS) and the regions of star formation (RSF) is found. Fraction of UGS, coincided in position with RSF, makes up $47\pm 8$\%. The coincided discrete gamma-ray sources possess, on average, harder energetic spectra and larger intensities with respect to the rest UGS. Annihilation of dark matter concentrating in RSF is supposed to account for a possible origin of gamma-radiation\footnote{Now a series of works exploring dark matter (DM) effects in star formation and evolution was issued \cite{DMstars}. In view of this we find worthwhile to make public given note about the revealed correlation between unidentified gamma-ray sources (UGS) and regions of star formation (RSF), reported at 29th Russian Conference on Cosmic Rays (3--7 August 2006) and published in ``Bulletin of the Russian Academy of Sciences: Physics'' {\bf 71} (2007), pp. 915-917.}.
\end{abstract}

%\newpage

\section{Introduction}

Investigations of cosmic gamma-radiation of high energy (50 MeV -- 5 GeV), carried out with special space stations COS-B (ESA), GRO-COMPTON (NASA), GAMMA (Roskosmos) in 1980--2000, discovered large amount of discrete sources of gamma-quanta of both galactic and extragalactic origin. Identification of gamma-ray sources with known objects (in radio, optic and other spectral ranges) is of great interest, since knowledge of optical, radio counterparts allows to ascertain the nature and the processes of generation of gamma-radiation. Significant part of gamma-ray sources are identified as yet. In Galaxy, these are young pulsars (neutron stars) and close binary star systems. Amongst extragalactic gamma-ray sources, these are near quasars, active galactic nuclei, blazars (variable objects with jets). Investigation of identified gamma-ray sources are very important for astrophysics of high energy. But unidentified gamma-ray sources (UGS), which are still not related to any known astrophysical objects, are likely to be of greater interest. They can make up a new class of active objects with a big energy release. The search for their counterparts in different spectral ranges can make clear their nature.

In the current paper an attempt to connect UGS of high energy ($E\ge 100$ MeV) with the regions of star formation (RSF) in our Galaxy has been undertaken.

\section{The used data}

For our purposes, data on coordinates and spectra were taken from catalogue \cite{gamma}, in which data were obtained at the space station gamma-telescopes during last years (COS-B, EGRET (station GRO-COMPTON)). Accuracy of the source coordinates, defined by tracks of $e^+e^-$-pairs in spark chambers, was $2\g - 3\g$ for $E_{\gamma}=100$ MeV. When discrete source is extracted from diffuse background, the error of source coordinate measurement was defined by statistics of the extracted gamma-quanta $N$, being, in essence, an error of a mean $\sigma\sim (1/N)^{1/2}\approx 0.5\g$. Angular resolution of gamma-telescopes was defined on the base of both calculations and their calibrating with beam of tagged gamma-quanta at accelerators. In the current work, we used in addition data on different expositions of the same source. It gave us additional information about coordinate accuracy taking into account the measuring errors in actual conditions of space experiment.

Comparison of UGS with RSF was carried out using the most complete and suitable catalogue of V.S.~Avedisova \cite{Avedisova}. This catalogue includes 3300 items of RFS and represents detailed data on their characteristics.

\section{Method of the search for coincidences of UGS and RSF}

\subsection{Optimization of accuracy of positional coincidences}

In the work devoted to identifying of gamma-sources with Wolf-Rayet stars \cite{WR-1,WR-2}, we inferred that actual accuracy of UGS coordinates makes up $\sim 0.5\g$. In the given work, UGS and RSF are considered to be coincident, if their angular coordinates hit into "error circle" with radius
$r=(\Delta b^2 + \Delta l^2)^{1/2} \le 0.3\g$,
where $\Delta b$ and $\Delta l$ are differences of latitudes and longitudes of the objects being compared. Thus, "error circle" had been diminished, and, respectively, number of accidental superpositions had decreased. Analysis has shown, that the chosen magnitude of $r$ were optimal for statistical confidence of coincidences between UGS and RSF.

\subsection{The choice of scanned region}

For the search for positional correlation between UGS and RSF, a band along galactic equator was chosen with $|b|\le 5\g$, $l=0\g\div 360\g$, where overwhelming number of RSF concentrates.

\section{The results of the search for coincidences}

\subsection{Correlations}

In the pointed band of latitudes $|b|\le 5\g$, there are $N_{\gamma}=40$ UGSs with energy $E>100$ MeV, from them $N_{\rm coinc}=19$ coincide in position with RSF: 12 are within $b=0\g\div 5\g$, 7 are within $b=0\g\div -5\g$. Information about "coincident" UGS and RSF (designation and galactic coordinates) are given in the Table.

\begin{table}
\caption{Data on coincident UGS and RSF}
\begin{tabular}{|p{1.2in}| p{0.6in} p{0.6in}| p{0.6in} p{0.6in}|p{0.7in}|} \hline
%\newline
\multicolumn{3}{|c|}{ UGS }& \multicolumn{2}{c|}{RSF} & \\ \hline
                 &   $l$   &   $b$   &  $l$    &   $b$   & $r$  \\ \hline  \hline
2CG 006+00       & 6.73    & $-0.14$ & 6.68    & $-0.25$ & 0.12  \\ \hline
2CG 075+00       & 75.46   &   0.60  & 75.45   &  0.73   & 0.15  \\ \hline
2CG 078+01       & 78.12   &   2.23  & 78.25   &  2.08   & 0.20  \\ \hline
2CG 121+04       & 121.00  &   3.98  & 121.18  &  4.02   & 0.18  \\ \hline
2CG 218$-01$     & 218.36  & $-0.53$ & 218.15  & $-0.57$ & 0.22  \\ \hline
2CG 284$-00$     & 284.45  & $-1.20$ & 284.24  & $-1.13$ & 0.23  \\ \hline
2CG 356+00       & 356.56  &   0.23  & 356.65  &  0.13   & 0.14  \\ \hline
2CG 359$-00$     & 359.53  & $-0.76$ & 359.44  &  0.01   & 0.27  \\ \hline
2EGJ 0618+2234   & 189.13  &   3.19  & 189.00  &  3.19   & 0.13  \\ \hline
2EGJ 1443$-6040$ & 316.28  & $-0.75$ & 316.16  & $-0.49$ & 0.29  \\ \hline
2EGJ 1718$-3310$ & 353.31  &   2.48  & 353.25  &  2.42   & 0.09  \\ \hline
2EGJ 1746$-2852$ & 0.17    & $-0.15$ & 0.18    & $-0.19$ & 0.05  \\ \hline
2EGJ 1825$-1307$ & 18.38   & $-0.43$ & 18.30   & $-0.39$ & 0.09  \\ \hline
2EGJ 1857+0118   & 34.80   & $-0.76$ & 34.76   & $-0.68$ & 0.09  \\ \hline
2EGJ 2033+4112   & 80.19   &   0.66  & 80.23   &  0.80   & 0.15  \\ \hline
2EGJ 2227+6122   & 106.60  &   3.14  & 106.90  &  3.16   & 0.30  \\ \hline
2EGJ 0852$-4343$ & 264.28  &   0.47  & 264.25  &  0.61   & 0.15  \\ \hline
2EGJ 1418$-6049$ & 313.31  &   0.29  & 313.45  &  0.18   & 0.18  \\ \hline
2EGJ 1903+0529   & 39.15   & $-0.08$ & 39.26   & $-0.05$ & 0.12  \\ \hline
\end{tabular}
\end{table}

The source 2CG284$-00$ coincides also with Wolf-Rayet star \cite{WR-1}. The rest 18 gamma-ray sources are identified for the first time.

\subsection{Probability of accident coincidence}

Two estimations have been made for probability of accident superposition of UGS and RSF in the scanned band.

A) Estimate following Gaussian distribution.

The probability of the coincidence is $\eta_{\rm coinc} = N_{\rm coinc}/N_{\gamma} = 0.47\pm 0.08$. From the $r$ given in the Table we get the mean "accuracy" of coincidence $\bar r =0.17\g$ and the area of coincidence $s_{\rm coinc}=\pi \bar r^2=0.09$ squared degrees. For the scanned area $S=10\g\times360\g=3600$ sq.deg.\ and the number of RSF inside scanned band $N_{\rm RSF}=3136$, probability of accidental coincidence is $\eta_{\rm accid}=N_{\rm RSF}\cdot s_{\rm coinc}/S=0.078$, what differs on 5 standard variances from the obtained probability of coincidence $\eta_{\rm coinc}$. According to normal distribution, probability of such accidental fluctuation makes up $w\approx 10^{-7}$.

B) Estimate following Poisson distribution.

The number of accidental coincidences between coordinates of UGS and RSF, as follows from data above, is $N_{\rm accid}=3.2$. The obtained number $N_{\rm coinc}=19$ exceeds significantly $N_{\rm accid}$. Probability of fluctuation, estimated following Poisson distribution, is $w=3.2^{19}e^{-3.2}/19!\approx 10^{-9}$.

From the obtained results one can conclude, that considerable number of UGS of high energy ($\sim 50\%$) are indeed connected with the regions of forming stars. Probability of accidental superposition of coordinates of the compared objects is extremely small.

\section{Analysis of the results}

\begin{figure}[!ht]
\includegraphics[scale=0.6]{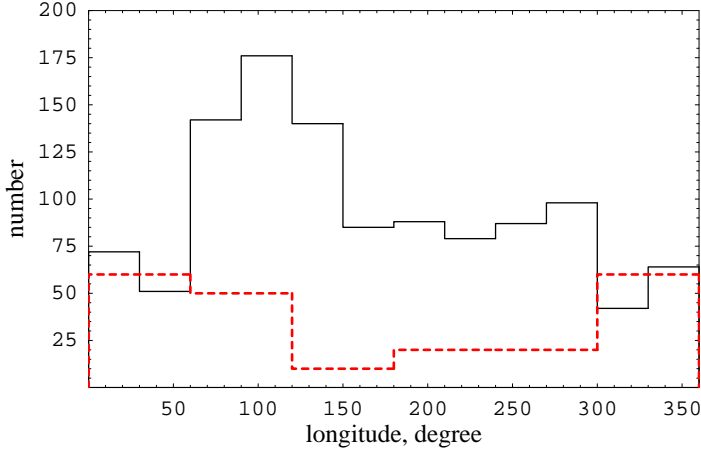}
\caption{Distribution in galactic longitude of RSF (solid line) and of coincident gamma-ray sources (dashed red line). Number of coincident gamma-ray sources is multiplied by 10.}
\label{distr_in_l}
\end{figure}

1. Distribution of RSF and "coincident" UGS in galactic longitude are shown on Fig.\ref{distr_in_l}. RSF are distributed inhomogeneously, there is a visible increase of RSF density within longitude interval $l=60\g\div 150\g$. Events of coincidence of UGS are approximately uniform in $l$, but their number is too small to compare in details with distribution of RSF. One can conclude, that given distributions does not confirm but does not contradict to actual origin of coincidences of UGS and RSF.

2. Observable characteristics of RSF has been analyzed. They are divided into the following groups: O (optical sources), M (molecular lines, maser sources), R (radio sources), IR (infrared sources). IR-objects compose the biggest part of RSF (76\%). Amongst RSF coincided with UGS, IR-objects are also dominant, their fraction is $80\pm 13$\%. So, analysis did not infer any preference of RSF connected with gamma-ray sources.

\begin{figure}[!ht]
\begin{tabular}{l  l}
\includegraphics[width=0.45\textwidth, height=0.3\textwidth]{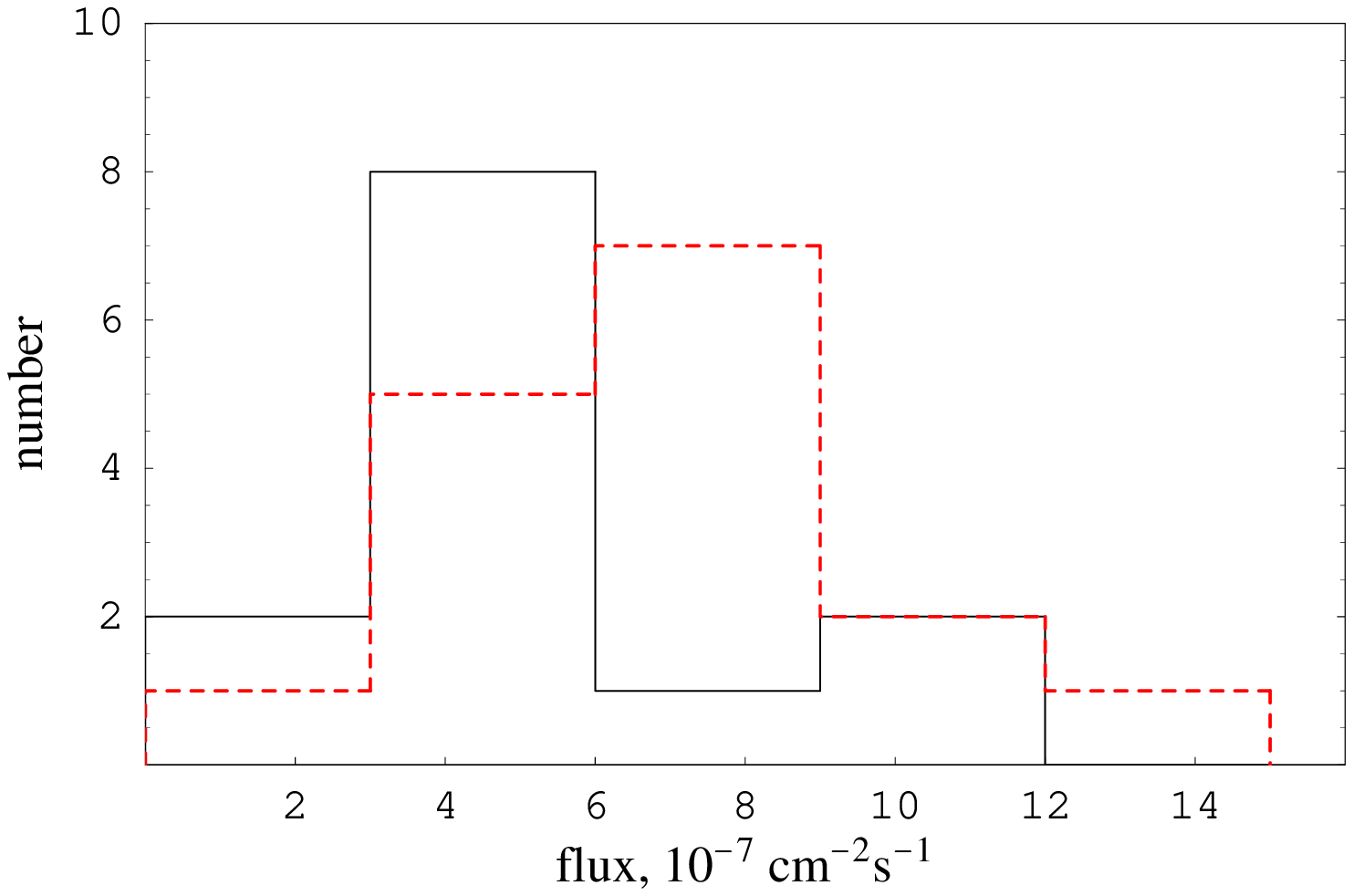} &
\includegraphics[width=0.45\textwidth, height=0.3\textwidth]{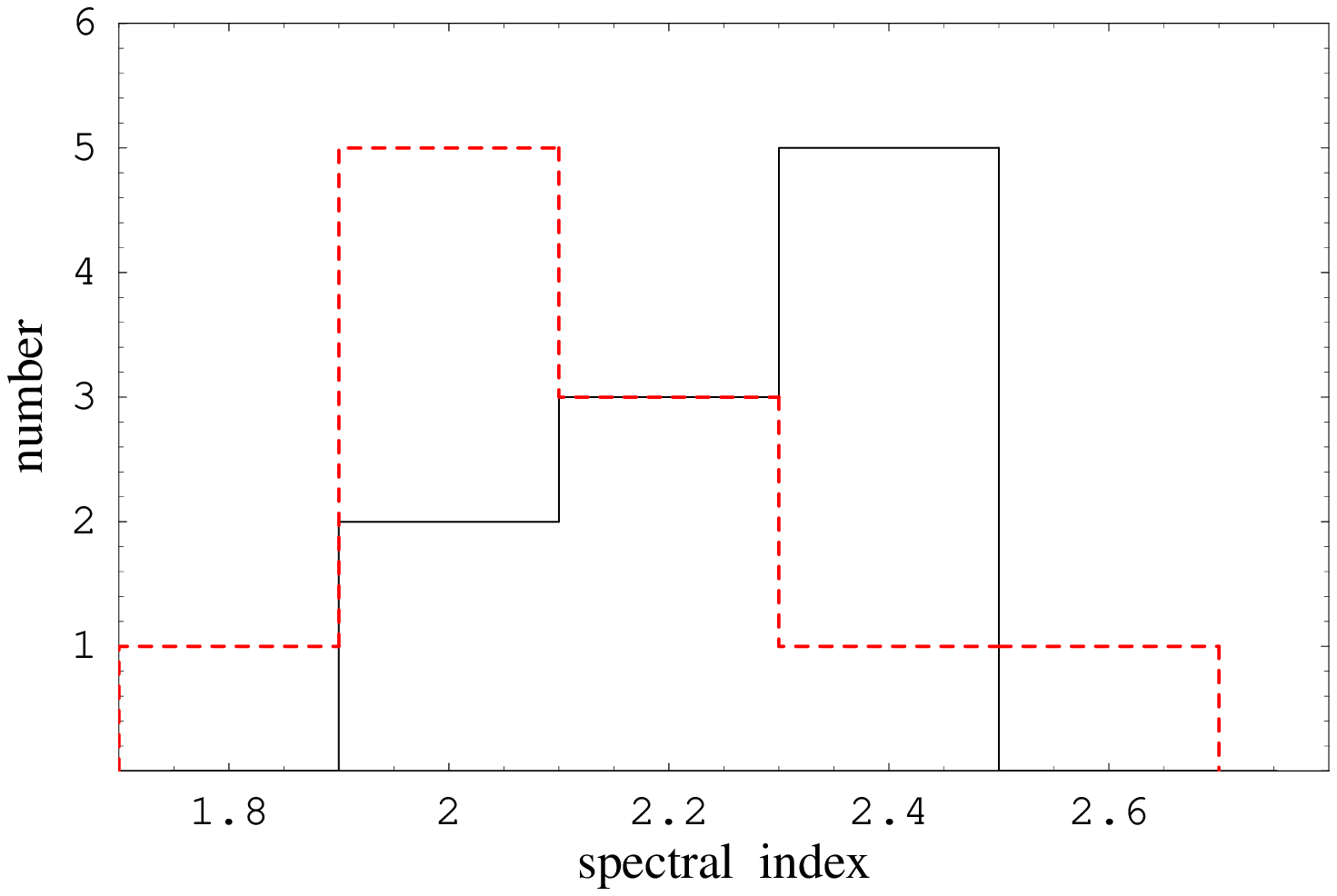}
\end{tabular}
\caption{Distribution in flux $F$ (left) and in spectral index $\alpha$ (right) for "coincided" (dashed red line) and "not coincided" (solid line) UGS within band $|b|\le 5\g$ }
\label{distr_in_F_al}
\end{figure}

3. Comparative analysis in gamma-ray fluxes $F$ and index of energetic spectra $\alpha$ has been carried out for UGS coincided and not coincided with RSF within scanned band. The results are shown on the Fig.\ref{distr_in_F_al}. As one can see, distribution of coincided and not coincided UGS overlap, but the mean values differ notably: $\alpha=2.0\pm 0.04$, $F=(6.1\pm 0.2)\times 10^{-7}$ cm$^{-2}$s$^{-1}$ for "coincided" UGS, $\alpha=2.2\pm 0.04$, $F=(4.4\pm 0.2)\times 10^{-7}$ cm$^{-2}$s$^{-1}$ for "not coincided" UGS. Deviations between them are considerable especially for fluxes $F$. "Coincided" gamma-ray sources have greater fluxes and harder energetic spectra, what can be regarded as a first indication to peculiarity of these objects, however being not very reliable because of small statistics and large measuring errors.

Nevertheless, on the base of obtained results we made one supposition about possible nature of given gamma-ray sources. It can be connected with a dark matter concentrating in RSF, annihilation of which leads to the enhanced flux of gamma-rays of high energy.

\section{Relation of RSF with dark matter}

Dark matter (DM) can be supposed to concentrate in regions of forming stars. One can pick out several mechanisms for such concentration. Baryonic and nonbaryonic components of matter can form density inhomogeneity together (either starting from the very beginning, or having merged subsequently). In this case baryons are situated inside a clump of DM. Also, it is possible that baryonic matter, being contracted, "draws together" DM due to adiabatic mechanism of energy loss by DM in varying gravitation field of baryons. Both mechanisms are likely to be most effective for large inhomogeneities (with mass $\gsim 10^6 M_{\odot}$) \cite{Khlopov}. In addition, density of DM should be enhanced owing to partial convergence of trajectories of DM particles to the attractive center created by baryons \cite{nv}. Besides of that, DM particles, going through baryonic matter, can "get stuck" in it and accumulate with time, efficiency of what is dependent from respective interaction properties. In reality, concentration of DM can come out from superposition of many mechanisms in different degree, where operation of one of them provides favourable conditions for another.

Annihilation rate of DM grows as its density squared, so annihilation source shows itself in region of highest density. Amongst annihilation products, produced in the result of cascade of transitions and decays of particles, there can be gamma-quanta with energy of hundreds MeV and above. Spectrum in this case should distinguish oneself with greater hardness.

\section{Conclusion}

The results of the undertaken search for coincidences of UGS with RSF in Galaxy point out at relation between RSF and UGS of high energy. Fraction of "coincided" UGS is big enough ($47\pm 8$\%), what gives the reason to consider the carried out identification to be meaningful for clarification of UGS nature. However statistics of UGS and number of coincidences with RSF is insufficient for decisive conclusion.

Before obtaining new observation data in forthcoming experiments (GLAST, AGILE, GALA and other) it is reasonable to carry out a detailed analysis of RSF, first of all with view of extraction of most massive and compact objects in the regions of star formation. Comparison of them with unidentified gamma-ray sources can give then more definite results.

\section{Acknowledgements}

Authors are grateful to V.S.~Avedisova for help in the work and analysis of results. The work is fulfilled in a framework of project of Rosobrazovanie RNP2.2.2.2.8248.

\end{document}